\documentstyle[12pt,fullpage,epsfig]{article}
\def\be{\begin{equation}}
\def\ee{\end{equation}}
\def\bea{\begin{eqnarray}}
\def\eea{\end{eqnarray}}

\def\gsim{\:\raisebox{-0.5ex}{$\stackrel{\textstyle>}{\sim}$}\:} 
\def\gluino{m_{\tilde g}}
\def\ul{ m_{{\tilde u}_L}}
\def\ur{ m_{{\tilde u}_R}}
\def\dl{ m_{{\tilde d}_L}}
\def\dr{ m_{{\tilde d}_R}}
\def\el{ m_{{\tilde e}_L}}
\def\er{ m_{{\tilde e}_R}}
\def\nuel{ m_{{\tilde \nu}_e}}
\def\tone{ m_{{\tilde t}_1}}
\def\ttwo{ m_{{\tilde t}_2}}
\def\bone{ m_{{\tilde b}_1}}
\def\btwo{ m_{{\tilde b}_2}}
\def\stauone{ m_{{\tilde \tau}_1}}
\def\stautwo{ m_{{\tilde \tau}_2}}
\def\snutau{ m_{{\tilde \nu}_\tau}}
\def\charone{m_{\tilde \chi^\pm_1}}
\def\chartwo{m_{\tilde \chi^\pm_2}}
\def\neutone{m_{\tilde \chi^0_1}}
\def\neuttwo{m_{\tilde \chi^0_2}}
\def\neutthree{m_{\tilde \chi^0_3}}
\def\neutfour{m_{\tilde \chi^0_4}}
\def\hsm{m_h}
\def\hlrg{m_H}
\def\ha{m_A}
\def\hpm{m_{H^\pm}}

\begin{document}
\begin{flushright}
TIFR/TH/00-42 \\
\end{flushright}
\bigskip
\begin{center}
{\large{ {\bf LHC Signature of the Minimal SUGRA Model with a Large Soft Scalar
Mass}}} \\[1cm]
Utpal Chattopadhyay$^1$, Amitava Datta$^2$, Anindya Datta$^3$, \\ 
Aseshkrishna Datta$^3$ and D.P. Roy$^1$ \\[1cm]
{\em $^1$Department of Theoretical Physics, Tata Institute of Fundamental
Research, Mumbai - 400 005, India \\ 
$^2$Physics Department, Visva Bharati, Santiniketan - 731 235, \\West Bengal, 
India \\ 
$^3$Mehta Research Institute, Jhusi, Allahabad - 211 019, India \\ 
}
\end{center}
\bigskip\bigskip

\begin{abstract}

Thanks to the focus point phenomenon, it is quite {\it natural} for the
minimal SUGRA model to have a large soft scalar mass $(m_0 > 1~{\rm
TeV})$.  A distinctive feature of this model is an inverted hierarchy,
where the lighter stop has a significantly smaller mass than the other
squarks and sleptons.  Consequently, the gluino is predicted to decay
dominantly via stop exchange into a channel containing $2b$ and $2W$
along with the LSP.  We exploit this feature to construct a robust 
signature for this model at the LHC in leptonic channels with 3-4 $b$
tags and a large missing-$E_T$.
\end{abstract}
 
\newpage

The minimal SUGRA model represents the most attractive model of low
energy supersymmetry in terms of simplicity and economy of 
parameters~\cite{nilles84}.  Besides it can naturally account for 
the electroweak symmetry
breaking as well as the suppression of flavour changing neutral
current effects.  The basic parameters of the model are $m_0$,
$M_{1/2}$, $A$, $B$ and $\mu$ -- i.e. the soft supersymmetry breaking
scalar and gaugino masses, the trilinear and bilinear couplings, along
with the supersymmetric Higgs mass parameter.  The last two can be
determined in terms of the two Higgs vacuum expectation values, $v_1$
and $v_2$, using the two minimisation conditions.  The first one
determines the $B$ parameter in term of
\be
v^2 = v^2_1 + v^2_2 = 2m^2_Z/(g^2 + g^{\prime 2}) \simeq 175~{\rm
GeV}, 
\label{one}
\ee
and the ratio $v_2/v_1 \equiv \tan\beta$.  The second condition gives 
\be
{1\over2} m^2_Z = {m^2_{H_1} - m^2_{H_2} \tan^2 \beta \over \tan^2
\beta - 1} - \mu^2 + \Delta_R, 
\label{two}
\ee
where the last term comes from the radiative correction to the Higgs
potential. 

Thus for any $\tan\beta$, the naturalness of the electroweak scale
requires $m^2_{H_2}$ and $\mu^2$ to be of the order of $m^2_Z$, so
that there is no large cancellation between these 
parameters~\cite{barbieri88,anderson95}.
Since $m^2_{H_2}$ is linearly related to the soft mass parameters
$m^2_0$ and $M^2_{1/2}$ via its RGE, one usually assumes the
naturalness criterion to imply $m_0$ and $M_{1/2} < 1~{\rm TeV}$
each.  Indeed most of the phenomenological works on the minimal SUGRA
model are based on this assumption.  It has been recently emphasised
in ref. ~\cite{feng9909334}, 
however, that for physical values of the top Yukawa and the
gauge couplings, $m^2_{H_2}$ turns out to be practically
independent of its GUT scale value $m^2_0$ for $\tan\beta \gsim 5$.
Moreover, contrary to some earlier apprehensions, a large value of
$m_0$ seems to lead to a cosmologically interesting dark matter
density~\cite{feng0004043}.  Besides a large $m_0$ would also 
alleviate the potential
conflict of the minimal SUGRA model with the electric dipole moments
of electron and neutron~\cite{kizukuri92}.  Thus the minimal SUGRA model with a
large soft scalar mass $(m_0 > 1~{\rm TeV})$ seems to be attractive
both on theoretical and phenomenological grounds.  We shall analyse
the signature of this model at the large hadron collider (LHC) by
exploiting the distinctive characteristics of the large $m_0$ limit.
Indeed we shall see that they lead to a more robust signal at LHC
compared to the canonical SUGRA model.
\bigskip

\noindent {\underbar{\bf The Model}}
\medskip

For qualitative understanding of the model it is instructive to look
at the approximate expressions for the electroweak scale scalar masses
in terms of the universal soft mass parameters at the GUT scale.  We
shall neglect the GUT scale $A$ parameter, which is unimportant for
the present consideration; and assume not too large $\tan\beta$ where
the $b$ Yukawa coupling is relatively less significant.  Then analytic
solutions to the one-loop RGE give
\bea
m^2_{H_2} &=& m^2_0 - {3 \over 2} y m^2_0 + O(M^2_{1/2}),
\nonumber \\[2mm] 
m^2_U &=& m^2_0 - y m^2_0 + O(M^2_{1/2}), \label{three} \\[2mm] 
m^2_Q &=& m^2_0 - {1\over 2} y m^2_0 + O(M^2_{1/2}), \nonumber 
\eea
where $U$ and $Q$ refer to the 3rd generation singlet and doublet
squarks~\cite{carena94}.  Here 
\be
y = {h^2_t \over h^2_f} = {1 + 1/\tan^2\beta \over 1 +
1/\tan^2\beta_f}, 
\label{four}
\ee
where the subscript $f$ denotes the fixed point values of the top
Yukawa coupling and the corresponding $\tan\beta$ at the electroweak
scale.  The numerical coefficients of $y$ in (\ref{three}) simply
reflect the corresponding coefficients of $h^2_t$ in the RG evolutions
of $m^2_{H_2}$, $m^2_U$ and $m^2_Q$~\cite{ibanez85}.  

The top Yukawa coupling is related to its running mass,
\be
h_t = m_t (M_t)/v \sin \beta,
\label{five}
\ee
which is in turn related to the physical top quark mass $M_t$ via
\be
M_t = m_t (M_t) \left[1 + \Delta_{\rm QCD} + \Delta_{\rm SUSY}\right].
\label{six}
\ee
The QCD and SUSY radiative corrections add about 6\% and 4\%
respectively to the running mass to arrive at the physical top pole
mass, $M_t = 175 \pm 5~{\rm GeV}$~\cite{revparticle}.  
It is well known now that a
physical top mass of 175 GeV corresponds to the fixed point value,
$\tan\beta_f \simeq 1.5$ at the electroweak scale~\cite{casas98}, 
which defines
the lower limit of $\tan\beta$ in this model.  Of course
such a low value of $\tan\beta$ is ruled out by the recent LEP
limit on the lightest Higgs boson $(h^0)$ mass~\cite{revparticle}, 
suggesting $\tan\beta > 2(4)$ for 
maximal (small) stop mixing.  Substituting this
value of $\tan\beta_f$ in (\ref{four}) gives 
\be
y = {1+1/\tan^2\beta \over 1.44} \simeq {2\over3} ~{\rm for}~
\tan\beta \gsim 5.
\label{seven}
\ee
Thus over a large range of $\tan\beta$, which is also 
favoured by the above mentioned LEP data, one has 
\be
m^2_{H_2} \simeq O(M^2_{1/2}),
\label{eight}
\ee
i.e. the $H_2$ mass at the electroweak scale is practically
independent of its GUT scale value $m_0$.  This is the so called focus
point phenomenon~\cite{feng9909334,feng0004043}.

The corresponding values of the 3rd generation squark masses are 
\bea
m^2_U &\simeq& {1\over3} m^2_0 + O(M^2_{1/2}), \label{nine}
\\[2mm] 
m^2_Q &\simeq& {2\over3} m^2_0 + O(M^2_{1/2}). \label{ten}
\eea
The remaining scalar masses are not driven by the top Yukawa coupling,
and hence satisfy
\be
m^2_R \simeq m^2_0 + O(M^2_{1/2}),
\label{eleven}
\ee
where the subscript $R$ represents $H_1$, sleptons and the squark of
the 1st two generations as well as the right-handed bottom squark
$\tilde b_R$.  Thus the minimal SUGRA model with large $m_0$ implies
an inverted mass hierarchy, where $\tilde t_R$ and to a lesser
extent $\tilde t_L,\tilde b_L$ are lighter than the remaining squarks
and sleptons, while all of them are expected to be heavier than the
gluino.  This in turn implies that gluino should decay dominantly via
the lighter stop exchange, i.e.
\be
\tilde g {\buildrel \tilde t_1 \over \longrightarrow} t\bar t
\tilde\chi^0_i, ~t\bar b \tilde\chi^\pm_i \rightarrow b\bar b WW\tilde
\chi^0_1 X,
\label{twelve}
\ee
where $X$ represents any other particles like $Z,h^0$ and $W^+W^-$,
which can result from the cascade decay.  Thus one expects that the
decay of a gluino pair, produced at the LHC, will give a distinctive
signal containing $4b$ quarks and $4W$ bosons along with a large
missing $E_T$ $(E\!\!\!/_T)$, carried away by the 2 LSPs.  Of course
some of the signal characteristics like the presence of multiple $b$ 
quarks are common to many SUSY models having inverted mass
hierarchy~\cite{baer9912494}.  However, we feel that the 
minimal SUGRA with large
$m_0$ represents by far the most well motivated and well defined model
of this kind. A distinctive characteristic of this model is the presence 
of multiple $W$ bosons in the signal along with 4$b$ quarks.  It may 
be noted here that the LHC signature of this
model has been recently considered in ref.~\cite{allanach2000}, 
which has not discussed however the distinctive characteristics of the model,
mentioned above. 
\bigskip

\noindent \underbar{\bf The SUSY Spectra}
\medskip

\nobreak
We have studied the radiative electroweak symmetry breaking and the
resulting fine-tuning parameter in the $m_0 - M_{1/2}$ plane to select
optimal values of these parameters and the resulting SUSY Spectra.
For this purpose we have modified the radiative electroweak symmetry
breaking code of ref.~\cite{chan98}, which uses two-loop RGE along with
two-loop QCD correction to the top quark mass (eq. \ref{six}), by
adding the one-loop SUSY correction to the latter~\cite{bagger97}.  
Indeed this
plays a very significant role in bringing down the focus point to the
electroweak scale.  The radiative correction to the Higgs potential in
eq. (\ref{two}) is evaluated using the complete one-loop 
result~\cite{oneloopeff}.  The
dominant contribution can be written in terms of the average stop mass
as 
\be
\Delta_R \simeq {3h^2_t \over 16\pi^2} m^2_{\tilde t} \left[{1\over2}
- \ell n\left({m_{\tilde t} \over Q}\right)\right].
\label{thirteen}
\ee
In order to keep this radiative correction small we shall use the
minimisation condition (\ref{two}) at a 
scale $Q = m^{\rm max}_R/2$~\cite{chan98}, which
is close to the mean stop mass (eqs. \ref{nine}-\ref{eleven}).  The
resulting $\mu^2$ is then evolved down to the electroweak scale, $Q
\simeq m_Z$, along with the other mass parameters.

The functional dependence of the electroweak scale on the SUSY
parameters is determined via eq. (\ref{two}).  The sensitivity of this
scale to fractional variations of these parameters is measured by the
sensitivity coefficients 
\be
c_a \equiv \left|{a \over m^2_Z} {\partial m^2_Z \over \partial
a}\right|,
\label{fourteen}
\ee
where $a$ represents $m_0,M_{1/2}$ and $\mu$\footnote{As mentioned
earlier we set the GUT scale $A$ parameter to zero, since it is
unimportant for our purpose.  Nonetheless this parameter is generated
at the electroweak scale by the RGE, resulting in $\tilde t_L - \tilde
t_R$ and $\tilde b_L - \tilde b_R$ mixing.}.  And the fine tuning
parameter is defined by the largest of these coefficients
\be
c = {\rm max}\{c_{m_0}, c_{M_{1/2}},c_\mu\}.
\label{fifteen}
\ee
\noindent
It should be noted here that there is an ongoing debate on whether
the sensitivity coefficients should include the variation of 
$m_Z^2$ with respect to $h_t$. On this issue we agree with the authors of 
ref.~\cite{feng9909334,feng0004043} that the SM parameters like
gauge and Yukawa couplings may have origins totally unrelated to 
SUSY. Hence, it would not be appropriate to include them in 
determining the {\it natural} ranges of the SUSY parameters.

Fig. 1 shows the contours of this fine-tuning parameter in the $m_0 -
M_{1/2}$ plane for $\tan\beta = 10$.  While the fine-tuning parameter
is seen to increase steadily with $M_{1/2}$, it is remarkably
insensitive to $m_0$.  In fact for a given $M_{1/2} \sim 500$ GeV, one
seems to require by far the least fine-tuning at $m_0 = 1500$ (2000)
GeV for $M_t = 175$ (180) GeV.  The figure also shows the
contours\footnote{To be definitive we have chosen the positive 
sign of $\mu$ throughout this work; but the results are not expected 
to be sensitive to this choice.} of
constant $\mu$.  The $\mu = 100$ GeV contour defines
the physical 
boundary, as it represents the lower limit of $\mu$ coming from
chargino search at LEP~\cite{revparticle}.  This of 
course lies close to the no
Electroweak Symmetry-breaking boundary, corresponding 
to $\mu^2 < 0$~\cite{falk99}.  It
should be noted that the value of $\mu$ and the resulting physical
boundary are very sensitive to the exact value of the top quark mass.
We have checked that these values agree quantitatively with those
obtained using the recent version of ISASUGRA~\cite{isasugra748}, 
with the same choice of scale, $Q = m^{\rm max}_R/2$.

We see from Fig. 1 that the largest values of $M_{1/2}$ and $m_0$,
consistent with a fine-tuning parameter $c < 250$ and the physical
boundary for $M_t = 175$ GeV, are
\be
M_{1/2} = 500~{\rm GeV}, ~~m_0 = 2000~{\rm GeV}.
\label{sixteen}
\ee
Therefore we choose this point for the computation of the SUSY
spectra.  Note that the fine tuning parameter at $m_0 = 2000$ GeV is
about the same as at $m_0 = 200 - 400$ GeV.  In view of the
sensitivity of $\mu$ to the exact value of the top mass, we have shown
in Table~1 two sets of SUSY spectra for $M_t = 175$ as well as 180 GeV,
leading to $\mu \ll M_{1/2}$ and $\mu \sim M_{1/2}$
respectively.  In order to facilitate comparison with other works
these SUSY spectra have been obtained by using the ISASUGRA 
code~\cite{isasugra748} with the above mentioned choice of scale.  The main  
difference between the two spectra is that one corresponds to a
higgsino LSP, while the other corresponds to a gaugino LSP as in the
canonical SUGRA model.  We shall analyse the LHC signature for both
the cases.  It may be added here that although we could have 
gone to a still larger $m_0$ for $M_t = 180$ GeV, it would make little
difference to the final signature. 
\bigskip

\noindent \underbar{\bf The LHC Signature}
\medskip

\nobreak
It is clear from the SUSY spectra of Table 1 that the dominant SUSY
signal at LHC is expected to come from gluino pair-production
\be
gg(q\bar q) \rightarrow \tilde g \tilde g.
\label{seventeen}
\ee
We have calculated this leading order cross-section using the CTEQ4M
parametrisation~\cite{lai97} at a QCD scale $Q = m_{\tilde g}$; and multiplied
it by a $K$ factor of 2 to account for the NLO effects~\cite{beenakker97}.

Since the gluino is expected to dacay dominantly via stop exchange,
one has to take account of the Yukawa couplings of top and bottom
squarks along with the gauge couplings.  The relevant formulae are
given ref.~\cite{bartl91}.  We have used them to calculate the branching ratios
of gluino decay.  The channels of eq. (\ref{twelve}) account for an
overall BR of 80\% for both cases.  But in the higgsino LSP case a
significant fraction of this $(BR \simeq 35\%)$ corresponds to one of
the $W$'s being off-shell due to the near degeneracy of $\chi^\pm_1$ and
$\chi^0_1$.  The resulting branching fractions of gluino-pair decay
into the final states of our interest are shown in Table 2 for both
the cases.  In the gaugino LSP case (b) the final states containing
$4b 4W 2 \chi^0_1 \cdots$ have a $BR \simeq 60\%$, while in the
higgsino LSP case (a) the final states containing $4b 3W 2 \chi^0_1
\cdots$ have a $BR \simeq 50\%$ after taking into account the smaller channels not shown in Table 2.  The presence of 3-4 on-shell $W$
bosons lead to distinctive signals in their leptonic decay channels,
accompanied by multiple $b$-tags and a large missing-$E_T$
$(E\!\!\!/_T)$, carried away by the LSP pair.  They provide a robust
signature for the model at the LHC, with very little SM background, as
we see below.

Using a parton level Monte Carlo routine we have computed the signal
cross-sections in the isolated 1-$l$, 2-$l$, same sign 2-$l$ and 
3-$l$ channels ($l = e,\mu$), accompanied by $\geq 3$ $b$-tags 
and $E\!\!\!/_T$.  We
have also computed the irreducible SM background in these channels from
\be
gg(q\bar q) \rightarrow t\bar t t\bar t 
\label{eighteen}
\ee
using the MADGRAPH program~\cite{stelzer94}, with $Q = M_t$ as the QCD scale.

We have tried to simulate detector resolution by gaussian smearing of
the jet and lepton energies~\cite{richter99}
\be
\Delta E_j/E_j = 0.6/\sqrt{E_j} + 0.03, ~\Delta E_l/E_l =
0.15/\sqrt{E_l} + 0.01.
\label{nineteen}
\ee
The basic selection cuts are
\be
p^j_T, p^l_T,E\!\!\!/_T > 30~{\rm GeV}, ~|\eta_j| < 3, ~|\eta_{l,b}| <
2.5,
\label{twenty}
\ee
along with the jet-separation and lepton-isolation cuts
\be
\Delta R_{jj}, \Delta R_{lj} > 0.4,
\label{twentyone}
\ee
where $(\Delta R)^2 = (\Delta \phi)^2 + (\Delta \eta)^2$.  We also
require $\geq 3$ $b$-tags, assuming a tagging efficiency $\epsilon_b =
0.5$ per $b$-jet~\cite{richter99}.  For the 1-$l$ channel 
we require at least two 
pairs of accompanying jets with $65~{\rm GeV} < m_{jj} < 95~{\rm GeV}$,
to simulate two accompanying $W$ bosons.  While having very little
effect on the signal or the background from (\ref{eighteen}), it
helps to suppress background processes like $t\bar t b\bar b$, which
would otherwise be 7-8 times larger than the signal.  For the same
reason we require at least one pair of accompanying jets with 65 GeV
$< m_{jj} < 95$ GeV in the 2-$l$ channel, but not in the 2-$l$ $SS$
(same-sign dilepton) or the 3-$l$ channels. 

Figs. 2 and 3 show the signal and the background cross-sections against the
accompanying $E\!\!\!/_T$ for the 1-$l$, 2-$l$, 2-$l$  $SS$ and 3-$l$
channels for $M_t = 175$ and 180 GeV respectively.  Thanks to the
accompanying LSP pair, the signal has a much 
harder $E\!\!\!/_T$ distribution than the background in each channel.
Thus the signal can be effectively separated from the background by an
accompanying ${E\!\!\!/}_T$ cut of 100-200 GeV.  Such an
${E\!\!\!/}_T$ cut also suppresses other background processes like
$t\bar t b\bar b$ and $Wb\bar b b\bar b$.  Therefore the viability of
the signature is primarily determined by the signal size.  Table 3
lists the signal cross-sections in the four leptonic channels with
$\geq 3$ $b$-tags and ${E\!\!\!/}_T > 100$ GeV.  Increasing the
accompanying ${E\!\!\!/}_T$ cut to 200 GeV reduces the signal
cross-section by only 15\%.  With the expected annual luminosity of
$100~{\rm fb}^{-1}$ at the high luminosity run of LHC, one expects to
see 30-50 signal events in the single-lepton and 10-20 events in the
dilepton channel, with very little SM background.  Thus one can
unambiguously probe the minimal SUGRA model upto $M_{1/2} = 500$ GeV
and the largest allowed value of $m_0$.
\medskip

\noindent \underbar{\bf Summary}
\medskip

We have considered a minimal SUGRA model with a large $m_0 = 2$ TeV,
which requires no larger fine-tuning than $m_0$ of a few hundred GeV.
It implies an inverted hierarchy, where the lighter stop has a
significantly smaller mass than the other squarks and sleptons.
Depending on the exact value of the top quark mass, the LSP can be
either gaugino-like or higgsino-like.  In either case the inverted
hierarchy ensures a robust signature at LHC in the isolated 1-$l$, 2-$l$,
same-sign 2-$l$ and 3-$l$ channels, accompanied by $\geq 3$ $b$-tags and
a large $E\!\!\!/_T$.  Using this signature one can unambiguously
probe the model upto $M_{1/2} = 500$ GeV and the largest possible
value of $m_0$.  An investigation of the signature of this model
at the Tevatron collider upgrades is currently under progress.  

We thank the organisers of WHEPP-6, where this project was started.
We also thank Manuel Drees for many helpful communications.  Amitava
Datta and D.P. Roy acknowledge partial financial support from BRNS project,
No. 37/4/97/R \& D II.

\newpage
\begin{table}
\begin{center}
\caption{
SUSY spectra in GeV at $m_0=$ 2 TeV, $M_{1/2}=$ 500 GeV, 
and $\tan \beta=$10, for $M_t=$ 175 and 180 GeV.
}
\begin{tabular}{ccc}
\hline
\hline
Mass & ${\rm M_t}=175$ GeV & ${\rm M_t}=180$ GeV \\
\hline
$\ul$ & 2230  & 2227 \\
$\ur$ & 2209 & 2206 \\
$\dl$ & 2231 & 2228 \\
$\dr$ & 2207 & 2204 \\
$\el$ & 2030 & 2030 \\
$\er$ & 2010 & 2010 \\
$\nuel$  & 2029 & 2028 \\
$\tone$  &1489  & 1424  \\
$\ttwo$  & 1910 & 1883 \\ 
$\bone$  & 1902 & 1875 \\
$\btwo$  & 2190 & 2188 \\ 
$\stauone$  &1993  & 1993 \\ 
$\stautwo$  & 2022 & 2022 \\ 
$\snutau$   &2021  & 2020 \\
$\gluino$   &1283  & 1279 \\
$\charone$  & 119 & 401 \\ 
$\chartwo$  & 441 & 567 \\
$\neutone$  & 105 & 215 \\ 
$\neuttwo$  & 134 & 402 \\
$\neutthree$ & 225 & 541 \\
$\neutfour$  & 441 & 568 \\
$\hsm$  & 118 & 121  \\
$\hlrg$ & 2009 & 2077 \\
$\ha$   & 2007  & 2075 \\
$\hpm$  & 2010 & 2078 \\
$\mu$   & 129  & 550 \\
\hline
\label{topcases}
\end{tabular}
\end{center}
\end{table}

\newpage
\begin{table}
\begin{center}
\caption{
Effective branching ratios for gluino pairs decaying into different
channels for $M_t$ = (a) 175 GeV and (b) 180 GeV, with tan$\beta=$10.
For (a) the generic final state is denoted by $4b + 2W + 2
\widetilde \chi_1^0 + X$, and for (b) it is $4b + 
4W + 2 \widetilde \chi_1^0 +X ^\prime$. Only those 
channels having branching ratios greater than 
1\% are shown.}
\begin{tabular}{|c|c|c|c|} 
\hline
\multicolumn{2}{|c}{$M_t$ = 175 GeV} 
& \multicolumn{2}{|c|}{$M_t$ = 180 GeV} \\ \hline \hline
X&Br.& X$^\prime$ & Br.\\ \hline \hline
$W W$ & 0.0420 &0 & 0.0412 \\ \hline
$W ^\ast W ^\ast$ & 0.1271 &$Z$ &0.0595 \\ \hline
$W ^\ast W$  & 0.1464 &$h$ &0.0914 \\ \hline  
$Z ^\ast W W$ & 0.0521 &$ZZ$ &0.0223 \\ \hline 
$Z ^\ast W ^\ast W$ & 0.0906 &$hh$ &0.0594 \\ \hline 
$Z ^\ast Z ^\ast W W$ & 0.0163 &$hZ$ &0.0752 \\ \hline 
$W ^\ast W W W$ & 0.0238 &$WW$ &0.0743 \\ \hline 
$W ^\ast Z ^\ast WWW$&0.0147 & $hhZ$ & 0.0165 \\ \hline
$W ^\ast W ^\ast WW$&0.0414& $WWh$ & 0.0823 \\ \hline
-&-&$WWZ$ & 0.0536 \\ \hline
-&-&$WWWW$ & 0.0334 \\ \hline
\end{tabular}
\end{center}
\end{table}   
\begin{table}
\begin{center}
\caption{
Signal cross-sections for the 1-$l$, 2-$l$, same sign 2-$l$, and 3-$l$ 
channels, accompanied by $\geq$ 3~b-tags and  
$E\!\!\!/_T >$ 100 GeV (tan$\beta=$10).
}
\begin{tabular}{|c|c|c|} 
\hline
{}& $M_t$ = 175 GeV  & $M_t$ = 180 GeV\\ 
{}& in fb & in fb \\ \hline
$\sigma _1$& 0.303 & 0.55   \\ \hline
$\sigma _2$& 0.094 &0.22  \\ \hline
$\sigma _2$(SS)& 0.047 & 0.1\\ \hline
$\sigma _3$ &0.013 & 0.06 \\ \hline
\end{tabular}
\end{center}
\end{table}   
\newpage
\begin{figure}[hbt]
\centerline{\epsfig{file=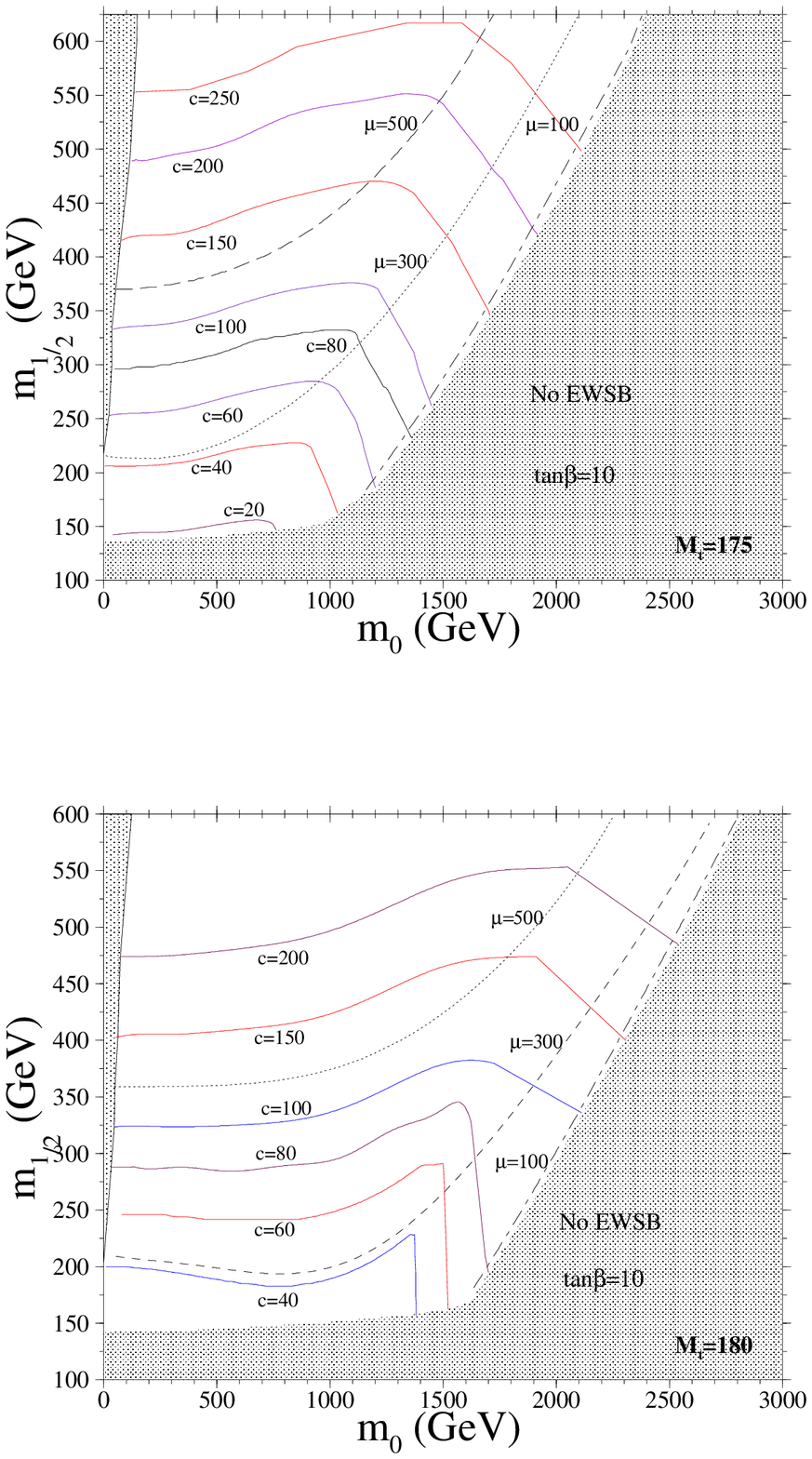,width=15cm}}
\vspace*{-1.5in}
\caption{
Fine tuning contours for $M_t=$175 and 180~GeV, 
with $\mu>0$. Contours of constant $\mu$ are also displayed. Shaded 
areas in the left sides are the charged LSP regions. 
}
\end{figure}

\newpage
\begin{figure}[hbt]
\centerline{\epsfig{file=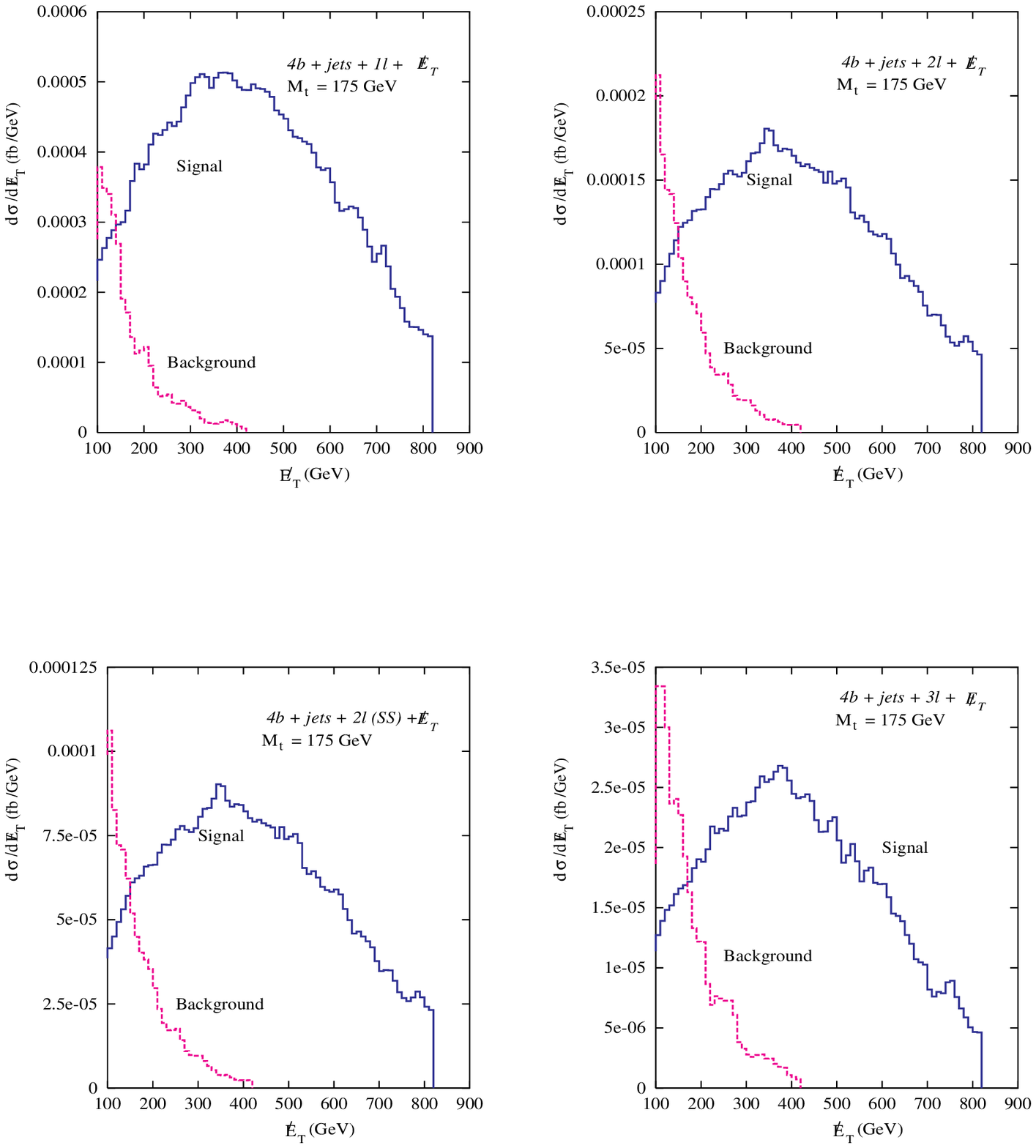,width=15cm}}
\vspace*{-1.5in}
\caption{
The SUSY signal and the irreducible SM background from 
$t \bar t t \bar t$ are shown against the accompanying 
$E\!\!\!/_T$ in the $1-l$, $2-l$, same sign $2-l$, and $3-l$ channels with
$\geq$~3~b-tags, for tan$\beta=$10 and $M_t=$175 GeV. 
}
\end{figure}

\newpage
\begin{figure}[hbt]
\centerline{\epsfig{file=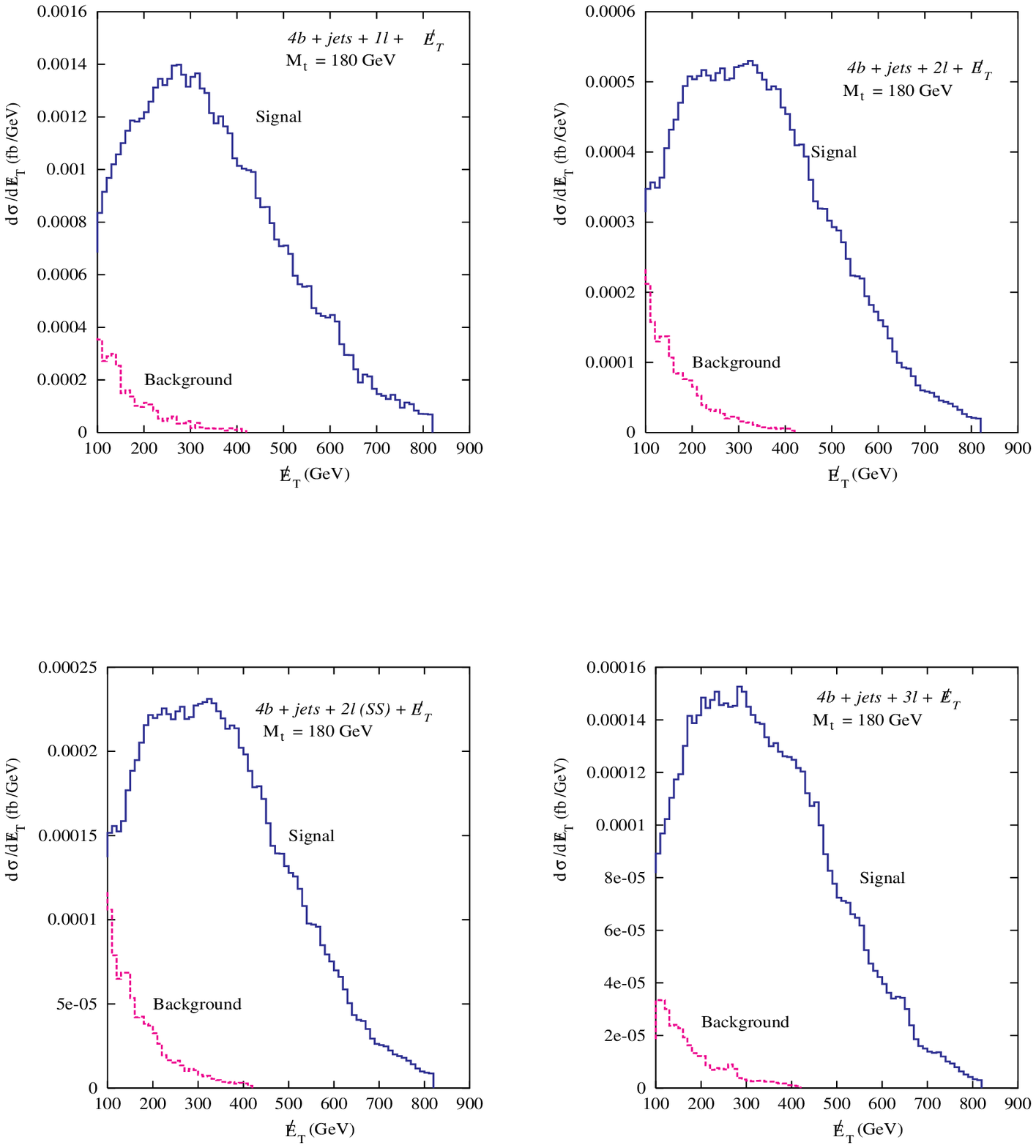,width=15cm}}
\vspace*{-1.5in}
\caption{
The SUSY signal and the irreducible SM background from 
$t \bar t t \bar t$ are shown against the accompanying 
$E\!\!\!/_T$ in the $1-l$, $2-l$, same sign $2-l$, and $3-l$ channels with
$\geq$~3~b-tags, for tan$\beta=$10 and $M_t=$180 GeV. 
}
\end{figure}

\end{document}